# Manifesto for Improved Foundations of Relational Model[1]

Witold LITWIN
Université Paris Dauphine, PSL

**Abstract.** *Normalized relations extended with inherited attributes can be more faithful to reality and support logical navigation free queries, properties available at present only through specific views. Adding inherited attributes can be nonetheless always less procedural than to define any such views. Present schemes should even typically suffice for relations with foreign keys. Implementing extended relations on popular DBSs appears also simple. Relational model should evolve accordingly, for benefit of likely millions of DBAs, clients, developers….*

## 1. Present Foundations

Relational database (DB) technology is the core for any type of modern databases: client, Web, cloud, Big Data… you name it. It appears nevertheless that the relational (data) model overlooked a fundamental issue from the beginning. Recall that we were all taught that for every normalized relation (table), e.g., in 3rd NF at least, adding an attribute can create a normalization anomaly. A canonical (Codd's) example was the so-called S-P database, with, essentially, the (base) relations SP (S#, P#, QTY), S (S#, SNAME, STATUS, CITY) and P (P#, PNAME, COLOR, WEIGHT, CITY). See the actual dataset of S-P in [1] or in a popular DB book, or there. Imagine now that one alters SP by adding, e.g., SNAME to SP with, for each SP.S# value, either the SNAME value for the same S.S# value or null, especially if SP.S# value is not in S as it can happen. Or, that, even, one creates such SP.SNAME instead of S.SNAME. Notice that we have in both cases the *functional dependency* FD : S# -> SNAME in SP, whenever SP.SNAME has a value.

For convenience, we may refer below to the original SP as to SP_ . Now, with SP.SNAME added, first, $2^{nd}$ NF of SP_ is gone. More importantly in practice, all the normalization anomalies follow. E.g., recall that we have SNAME = Smith for S# = S1 in S in the original S-P dataset. Hence, we now should insert Smith into every supply by S1 everywhere in SP, i.e. six times actually. This is already a hassle, compared to a single insert of Smith into S. In theoretical words, these inserts are necessary to maintain the functional dependency (FD): S# -> SNAME that appeared in SP. But, one could inadvertently insert into SP a supply by S1 with erroneous SNAME, say Smit. This would violate the FD and create obvious trouble for queries.

As known as well, altogether, the phenomenon would create an *insert* anomaly for SP. Similar nasty phenomena would create *update* and *delete* anomalies. See the literature for countless examples. The *normalization theory*, the basic piece of the relational model as well-known, concludes that one should avoid such side-effects. Hence, it claims for our case, that SNAME should be in S only. Same for every other attribute of S or P, except for the keys S# and P#. That is why S-P is as above. All the current teaching, textbooks & practice follow these foundations.

## 2. Proposed Foundations

The overlooked issue is an implicit assumption in the claim and all the normalization theory. Namely, that every attribute under consideration is a stored (base, non-calculable…) one (SA). Every attribute in S-P is so in particular. This assumption seemed apparently obvious to everyone involved. Namely, to Codd himself & to the cohort of research followers, students, practitioners… Actually, as our & some other work showed, most recently in [1], it is not. An added attribute could be an *inherited* one (IA), i.e., calculated through a relational expression with, perhaps, value expressions within, as in a view. Mathematically, the same relation would result as if we added an SA with always the same values as the IA. But, an IA cannot create a normalization anomaly. We called *Stored and Inherited Relation* (SIR) a stored relation extended with IAs. Here is how one can define in SQL SIR SP with SNAME being the IA we aim upon. Let us denote every IA as *Italic* from now on. E.g., the attributes of our SIR SP would be SP (S#, *SNAME*, P#,QTY).

For every tuple of SIR SP the value of *SNAME*, if any, should be that implied by the above discussed FD S# -> SNAME in SP (S#, SNAME, P#, QTY). E.g., we should have *SNAME* = Smith for every SP tuple with S# = S1. Observe that the following view SP has this property. It defines mathematically the same SQL relation as the stored SP (S#, SNAME, P#,QTY) would be. The original (normalized) SP had to be renamed for Create View SP, e.g., to the already mentioned SP_. No two relations in a DB can indeed share a proper name.

(1) Create View SP As (Select S#, SNAME, P#, QTY From SP_ Left Join S On (SP_.S# = S.S#));

The outer join is necessary since SP can have S# values not in S, as already mentioned. Within (1), one may see *SNAME* as defined through the (relational) *inheritance* expression (IE):

(2)  SNAME From SP_ Left Join S On (SP_.S# = S.S#);

To create SIR SP scheme, one can accordingly reuse the original Create Table SP with *SNAME* defined by (2) added to as it was within (1). Namely, suppose that the original statement creating the SP was:

(3) Create Table SP (S# Char 5, P# Char 5, QTY INT Primary Key (S#, P#));

Suppose also that SP_ below renames again this SP, but, this time, by default within SIR SP being created. Then, SIR SP could be declared through:

(4) Create Table SP (S# Char 5, SNAME, P# Char 5, QTY INT From SP_ Left Join S On (SP_.S# = S.S#) Primary Key (S#, P#));

As view SP (1), SIR SP (4) is mathematically the same SQL relation as would be SP (S#, SNAME, P#,QTY). The difference is that every SNAME value stored in the latter becomes an inherited one of *SNAME* in (4). The former inherits the data type from S as

---

well, we recall. To insert a new supply, it suffices to define SAs in (4) only. In other words, there is no need to set SNAME value anymore. E.g., suppose that one issues Insert SP (S# = S1, P# = P20, QTY = 100). *SNAME* = Smith will appear magically and always correctly to every query selecting SNAME in any tuple with SP.S# = S1. E.g., it will appear so to: Select * From SP Where S# = S1 And P# = P20;. It is the calculus through IE (2) in (4) that (virtually) propagates it. FD S# -> *SNAME* in SP cannot get violated anymore. The insert anomaly cannot occur accordingly as well. Likewise, SIR SP is free from the other anomalies.

As said, similar normalization issues concern every other attribute of S except for key S#. Likewise, it concerns every attribute of P, besides P#. Create Table with IE avoiding the whole trouble through adequate SIR SP could therefore be:

(5) Create Table SP (S# Char 5, SNAME, STATUS, S.CITY, P# Char 5, PNAME, COLOR, WEIGHT, P.CITY, QTY INT From SP_ Left Join S On (SP_.S# = S.S#) LEFT JOIN P On (SP_.P# = P.P#) Primary Key (S#, P#));

Here, IAs are all those without data type. As in (4), SP_ in (5) continues to name by default all the SAs in SP, constituting again the original SP. The underlying SQL view SP generalizing (1) and the basis for IE in (5) is:

(6) Create View SP AS (SP_.S#, SNAME, STATUS, S.CITY, SP_.P#, PNAME, COLOR, WEIGHT, P.CITY, QTY From SP_ Left Join S On (SP_.S# = S.S#) LEFT JOIN P On (SP_.P# = P.P#));

See [1] for the actual SA and IA values in SIR SP (5), given the 'biblical' S-P dataset. Observe also that IE in (5) is *less procedural* than (6), i.e., requires fewer characters, by the simplest measure of counting the characters necessary per statement. Recall that reducing procedurality was always among main goals of the database science, leading in particular to the replacement of the Codasyl and hierarchical DBs with the relational ones. Through SQL clauses specific to SIRs, i.e., beyond any SQL dialect at present, proposed in [1], IE in (6) could be even less procedural. E.g., observe first that SP.S# and SP.P# are foreign keys, in the sense of representing by key values tuples of other relations, S and P here, usually called *referenced* ones and, also as usual, where every referenced key bears the same (proper) name as the foreign key, as in S-P actually. Next, consider that for every foreign key R.A, referencing key C of some (referenced) relation F, A being perhaps composed, (A) implicitly denotes in R with respect to A, also all the other attributes of F except for F.C, in their original SQL order with respect to C. Then, the IE, hence the whole Create Table SP statement (5), could become shorter:

(7) Create Table SP ((S# Char 5), (P# Char 5), QTY INT From SP_ Left Join S On (SP_.S# = S.S#) LEFT JOIN P On (SP_.P# = P.P#) Primary Key (S#, P#));

Observe furthermore that if the foreign keys defined as just discussed are the only source for the IAs, making the outer joins implicit is easy. Our Create Table SP, defining all the IAs still as in Create View SP (6), we recall, could be then almost as non-procedural as the current Create Table SP for the original base SP only, namely becoming only:

(8) Create Table SP ((S# Char 5), (P# Char 5), QTY INT Primary Key (S#, P#));

Observe lastly that one can invert the "parentheses" rule with the usual notation for a relation scheme. Namely, a usually defined foreign key A would mean (A) and vice versa. E.g., the usually defined SP (3) would mean (8), while (8) would create the usual SP. Usual Create Table R defines then some SIR R without any added procedurality finally, i.e., by default, with respect to that creating the usual stored relation R. E.g., Create Table SP (3) would create our SIR SP (5) instead of the "biblical" SP.

Observe however that, while becoming the least procedural, such Create Table loses its backward compatibility. E.g., a recreation of the legacy "biblical" SP would create a different relation. Such outcome may affect legacy applications. For some time, there will be thus the trade-off between both versions of Create Table: more procedurality versus backward compatibility or vice versa. We guess the final say should go to DBAs.

Recall finally that the definitions of 2NF, 3NF… were proposed as they are to avoid anomalies in the relations concerned that would appear necessarily otherwise. E.g., every relation in 2NF is free of anomalies that would appear for a relation in 1NF only necessarily, as we discussed for SNAME. But all these definition addressed implicitly SAs only, as we pointed out. To remain equally discriminative in possible presence of IAs, all these definitions should be amended. They should explicitly address all the SAs of the relation of concern only. E.g., a relation should be in 2NF iff the projection on all its SAs only is in 2NF. Accordingly, the original SP with SA SNAME added to would continue to not be in 2NF and present the anomalies. In contrast, SP with IA SNAME instead, would be qualified as in 2NF and free of the discussed anomalies, as we have seen.

## 3. Why New Foundations?

In sum, the proposed foundations provide for base relations only with SAs that would not create the normalization anomalies for present foundations, extensible however at will with IAs. IAs cannot create the anomalies, but each would if it was an SA, as it was tacitly supposed. E.g., it would be so for every IA in SP defined by (5), (7) or (8). But, what practical advantages such IAs could bring to base relations? We see two major ones, see [1] for more.

   a.   A normalized base relation with IAs becomes a more faithful conceptual model.
   b.   IAs avoid the logical navigation necessary for many if not most of queries to the relation.

Indeed with respect to (a), observe that about every actual supply includes as conceptual properties (attributes) at least the names of the supplier and of the part. Thus every actual supply modelled by SP should include at least SNAME and PNAME, together, perhaps, with any other S-P attribute of a supplier or a part. None of these is in (the original) SP. That one is the poorest possible conceptual model of a supply, due to the normalization constraints, as widely known. SP with all the IAs defined in (5) is free of this drawback. Actually, it is the richest model of a supply within S-P.

With respect to (b), a query *Q* selecting any SA and any IA of SIR SP defined by (5), (7) or (8) for instance, could be free of the logical navigation. In contrast, every *Q'* equivalent to *Q*, i.e., defining the same outcome, issued towards the traditional SP would require the navigation. It necessarily addresses indeed several relations. The navigation consists then usually of inner or outer equijoin clauses over foreign and referenced keys, we recall. E.g., for our SIR SP, we could have *Q* : Selects S#, SNAME, QTY From SP. Then, *Q'* would require the logical navigation between (traditional) SP and S through the outer joins in (5). That navigation must make every *Q'* substantially more procedural than *Q*. The drawback obviously results from the poverty of SP conceptual model. The discussed properties clearly generalize to every *Q* towards every base SIR R with some best normalized SAs and IAs and every *Q'* towards "traditional" base R, i.e., with these SAs only.

The present practice for (a) and (b), if of importance for clients of a base relation, say R, is to provide a specific view R instead. At least in SQL, one has to rename R then, say to R_ again. Every such view R defines mathematically the same relation as R_ would be if it had also as SAs some or, more typically, all of its conceptual attributes that had to land in other relations, to get R_ normalized. For instance our views SP (1) and (6) could be specific views SP of SP_ with S and P being the other relations mentioned. Every such view R clearly provides for (a) and (b). Actually, it provides for these goals exactly as every SIR R created as above would do. E.g., it is obviously so for our view SP (1) and SIR SP (4), and for view SP (6) and SIR SP (5). Sometimes, a different specific view R is possible, [1]. Every such view has the same proper attribute names and the same tuples. A source (relation) name of a unique proper name in the latter view may however differ from the one in former view. As known, SQL queries do not need source names for unique attribute names. Every practical SQL query to former view is thus valid for the latter one and vice versa. Create View for the latter view may however, in addition, be substantially less procedural. This can make the latter view preferred.

The already mentioned specific to SIRs clauses guarantee nevertheless that for every specific view R, there is always SIR R with IE less procedural than Create View R, while providing for the same practical queries. E.g., for our canonical S-P example, by our measure, the procedurality, say *p*, of (6) is *p* = 150. The one of IE in (8) is *p* = 4. This IE is thus 37.5 times less procedural for the same result. By the other token, to get (a) and (b), requires only mere 4 more characters in (7), i.e. for *p* = 69 already of Create Table SP otherwise. If we invert the "parentheses" rule as we spoke about, then we save entire 150 characters at zero procedural cost with respect to the present SP scheme.

Likewise, for every SIR R, altering any specific view R should be always more procedural then altering the IE. By the same token, if altering R_ implies altering view R, the whole action is also more procedural than the equivalent altering of SIR R, possibly several times, [1]. By the same token, dropping SIR R should be always less procedural than dropping R_ and view R. All this is in particular true for our SIR SP with respect to any view SP, as above hinted and detailed in [1]. If thus not only the normalization, but also (a) and (b) are of importance for some relation R, as it was for our SP, (and should always be, as many believe already for decades), then, whatever is view R, there is always a more advantageous SIR R. In particular, if we inverted the "parentheses" rule as we spoke about, then, also, every present DB scheme with relations with foreign keys, e.g., S-P scheme, would apply as is on a SIR-enabled DBS like those sketched in next section. All the discussed benefits would result in consequence. Incidentally, imagine the effort of students, DBAs, clients, developers… on the logical navigation in queries or creation and maintenance of the specific views that would get saved if these findings had seen the lights when the relational model got proposed.

**4. Implementing SIRs on Popular DBSs**

The practice of our proposal should obviously start with an implementation on a popular DBS. For various reasons, MySQL seems the best starting point, along the gross architecture defined in [1]. That one manages SIRs through a new component termed *SIR-layer*. An internal representation of SIR R within the DBS underneath, can be (1) the stored relation R_ formed by all and only SAs of SIR R and (2) view R defining the same SQL relation as SIR R. E.g., SIR SP would become SP_ and view SP (6). SIR-layer typically directs then queries to SIR R towards view R. Most queries to SIR-layer with SP created using (5), (7) or (8) would address thus view SP (6). A few cases require specific processing. But, altogether, an implementation as outlined appears simple and with negligible storage and processing overhead at popular DBSs.

**5. Conclusion**

S-P was the mold for about every relational DB. Our examples and the benefits of the IAs extend accordingly and can be expected typical. The **proposed foundations should enter the DB teaching, textbooks & everyday practice.** It should become the basic knowledge that a normalized base relation, e.g., the "biblical" SP, we referred to also as to SP_, we recall, may still have even all its other actual conceptual attributes added to: SNAME,…, PNAME,…, without presenting any anomalies. Such attributes can be in particular functionally dependent on a part of a primary key, e.g., SNAME on S#. Hence, to say or write that adding an attribute with that property to a relation in 2NF or higher must create an anomaly is from now on simply untrue. To remain equally discriminative as till now, the definitions of 2NF, 3NF… in the current DB literature and, e.g., Wikipedia, should be rephrased accordingly, as we discussed.

It should become common awareness also that every normalized relation extended with IAs, as we discussed for SP, benefits from logical navigation free queries, impossible for that relation otherwise, i.e., for our SP_. Likewise, it should be the common knowledge that to create or alter the IAs in a base relation may be always made less procedural than to create or alter any view avoiding that navigation. E.g., as creating or altering IAs in (5), (7) or (8) was, with respect to view SP (6). In particular, it should become widely known that the current SQL definitions of relations with foreign keys can serve to create or alter such SIRs without any additional procedurality, i.e., simply as they are, as we hinted to for SP. Implementing SIRs on a popular DBS appears finally simple and with negligible overhead. We believe our proposal welcome therefore "better late than never", as the old saying goes, for benefit of likely millions of DBAs, clients, developers… of relational DBs.